\documentclass[10pt,letterpaper]{article}
\usepackage[top=0.85in,footskip=0.75in]{geometry}

\usepackage{amsmath,amssymb}

\usepackage{changepage}

\usepackage{textcomp,marvosym}

\usepackage{cite}

\usepackage{nameref,hyperref}

\usepackage[nopatch=eqnum]{microtype}
\DisableLigatures[f]{encoding = *, family = * }

\usepackage[table]{xcolor}

\usepackage{array}

\newcolumntype{+}{!{\vrule width 2pt}}

\newlength\savedwidth

\raggedright
\usepackage[aboveskip=1pt,labelfont=bf,labelsep=period,justification=raggedright,singlelinecheck=off]{caption}

\makeatletter
\renewcommand{\@biblabel}[1]{\quad#1.}
\makeatother

\usepackage{lastpage,fancyhdr,graphicx}
\usepackage{epstopdf}
\begin{document}
\vspace*{0.2in}

{\Large
\textbf\newline{Online word-of-mouth in West Africa: the effects of snowball sampling on completion rate, respondent demographics, and survey responses} 
}
\newline
\\
Alexander Zaitzeff\textsuperscript{1*},
Samuel Blazek\textsuperscript{1}
\\
\bigskip
\textbf{1} Two Six Technologies, Arlington, Virginia, United States of America
\\
\bigskip

%
%

*alexander.zaitzeff@twosixtech.com

\section*{Abstract}
We place geo-targeted advertisements on Facebook to encourage users to fill out an online survey, following a process known as river sampling. We discovered a large number and variety of users also came to our survey through snowball sampling, including shared social media posts and other word-of-mouth referral methods. In this article, we analyze the differences between the respondents from river and snowball sampling. We present evidence that the respondents obtained by snowball sampling are more likely to complete the survey and contain a higher fraction of new users and women than those obtained by river sampling. Additionally, the evidence indicates that users from snowball sampling give shorter responses and take less time on the survey than users from river sampling. We hope these findings provide insight for other researchers who incorporate social media strategies when fielding surveys.

\section*{Introduction}

In this study, we sought individuals in Zamfara state, Nigeria to complete a survey about current and anticipated future food prices. These surveys were fielded in support of broader research goals focused on examining the utility of local population-based knowledge systems in forecasting social and economic indicators. To solicit responses, we ran six five-day advertising campaigns over the course of seven months using Facebook and Instagram, social media platforms run by the company Meta. In these campaigns, we presented paid advertisements to social media users in Zamfara that sent them to the survey. We offered an incentive to respondents who completed the survey. In addition to this intentional sampling process, a snowball sampling process emerged organically. Specifically, social media users shared survey links on their timelines, thus appearing on their friends' and followers' feeds and inboxes, as well as posting to groups and sharing via direct messages. In these cases, new respondents came to the survey from word-of-mouth referrals through their social networks. In this article, we analyze how this additional organic sampling process affected our results.

\subsection*{River, snowball, and network sampling}
Increasing attention has been paid to non-probability survey sampling strategies~\cite{aapor2013report}, particularly as response rates to more traditional phone- and mail-based sampling strategies decline and risk increasing levels of potentially nonrandom attrition~\cite{stedman2019end}. 

River sampling and snowball sampling are both non-probability sampling techniques, and feature a number of similarities and differences. River sampling refers to recruitment of participants via an advertisement or other public posting, wherein some proportion of individuals who are exposed to the advertisement decide to engage with the advertisement and participate in the associated activity (for example, completing a linked web survey). In particular, river sampling over social media can offer access to a large number of respondents in a study geography, particularly as social media advertising systems are now able to feature advertisements not only on their own platforms, but on web sites that are integrated with their advertising ecosystems. However, coverage is still limited to the individuals who have accounts on the social media platforms in question. Moreover, similar to other channels used for survey recruitment including phone- and mail-based survey strategies, participants must ultimately make a decision to engage with the content presented to them. In the digital advertising domain, this decision is also referred to as \textit{conversion}. The factors influencing conversion can be difficult to estimate and may be an additional source of nonrandom participation (or lack thereof).

 Snowball sampling is considered synonymous with (chain) referral sampling in our study, and can be defined as leveraging word-of-mouth sharing by participants as the primary form of new participant recruitment~\cite{biernacki1981snow}. {Biernacki and Waldorf~\cite{biernacki1981snow}  provide one of the earliest detailed examinations of referral sampling and note: 
 \begin{quote}``the existing methodological literature that the chain referral method of sampling is a self-contained and self-propelled phenomenon, in that once it is started it somehow magically proceeds on its own. This, however, is simply not the case; rather, the researcher must actively and deliberately develop and control the sample's initiation, progress, and termination.''\end{quote} 
 However, in our design, unlike in traditional snowball sampling, we find that many of the controls available to researchers during in-person survey enumeration are not available to us. These include vetting the authenticity of responses via face-to-face or telephone-based validation and carefully selecting the initial respondents, known as ``seeds'' in respondent-driven sampling, to begin the referral process in a manner that minimizes selection bias and coverage bias -- two risks associated with chain referral sampling, due to oversampling dense networks and undersampling isolated members of a target population. Recruitment via online channels also induces new coverage biases, because members of a target population who are not on the selected online channels are not reachable. This is exacerbated if the selection of initial seeds for referral may feature selection or sampling biases, as in the study by Apuke et al~\cite{apuke2024information} which recruited initial ``seed'' respondents from the researchers’ social and personal networks.}
 
{Risks and biases that are relevant to snowball sampling have been evaluated in prior literature. The most relevant risks identified in~\cite{biernacki1981snow} to our study are verifying the eligibility of respondents and controlling the types, pacing, and number of referrals by any given participant. Additionally, there are some biases based on the Total Survey Error (TSE) framework~\cite{groves2010total} that apply. Naturally, sampling error is expected with a non-probability sampling method. Coverage error may arise based on the propagation channels available and the connectedness of the target population. Snowball sampling is often employed for qualitative methods that do not seek to build a large sample, so sample size issues can arise depending on design- and referral-specific factors. Snowball sampling may have different response rates and participation incentive structures owing to the interpersonal referral process, leading to systematic non-response biases. Adjustment and estimation errors are both significant risk factors if a non-probability sample with unknown biases is adjusted using statistical techniques to approximate a representative sample; for instance, by raking or weighting based on external demographic data.}
 
{Despite all of this, in many geographies around the world, overwhelmingly large percentages of the adult population are accessible via digital advertising on social media and other online channels. Moreover, although recruitment using advertisements is a stochastic process, it offers new mitigation opportunities to address validity threats compared to other methods of recruiting initial seeds in chain referral designs. In particular, mixed-methods approaches can mitigate the ``dense network'' effect by recruiting from more socially, geographically, or demographically diverse pools of initial respondents. For example, access to audiences across a large and prolific digital channel may offer advantages over time-location sampling in cases where an online population may feature forms of heterogeneity that a sample constrained by geographic and time factors would be unlikely to reflect. For this reason, results from surveys enumerated using an online strategy offer researchers a new opportunity for triangulation via insights into sampling or coverage issues that remain undiscovered in other chain referral designs.} 

Of additional note, variants of snowball sampling are often used to recruit hard-to-reach populations, based on the hypothesis that individual members of these populations often know other members, and that carefully controlled chain referral processes can thus propagate across the personal networks of members of a key population under study. {Online methods offer new mechanisms for reaching hard-to-reach populations that have been underexamined in prior literature.}
 
 {A closely related technique is network sampling. Past work on network sampling has leveraged respondent-driven sampling, which formalizes network sampling by attentively selecting seed respondents, tracking and defining successive samples in terms of referral waves, and measuring the proportion of saturation in terms of duplicates over time, to reach an ``equilibrium'' and thereby derive a sample that proponents argue is statistically independent from the original sample and approximates a probability sample of the frame~\cite{heckathorn2002rds}.}

\subsection*{Online sampling in Africa}
There has been widespread use of social media advertising in Africa to distribute surveys~\cite{freihardt2020can,olamijuwon2021characterizing,pham2019online,rosenzweig2020survey,wagenaar2012hiv}. This is due in part to the much lower cost of fielding online surveys and more immediate access to populations compared to traditional survey methods. Given these advantages and with continually increasing rates of connectivity across Africa~\cite{ezeoha2020africa}, research has particularly focused on assessing the suitability of social media-based sampling strategies for producing generalizable inferences and insights, including how using social media recruitment strategies compare to traditional survey methods in terms of coverage and response quality.

In~\cite{pham2019online}, the authors assess the efficacy of Meta's advertising platform in Kenya. They first compare Meta's audience estimates with 2009 Census data. They find that ``Facebook audience sizes generally reflect underlying population density.'' However, they note that remote regions are underrepresented on Facebook. They proceed to use Meta's audience targeting tools to run a survey and compare the results to traditional surveys run in Kenya. Compared to the more traditional survey method, their survey skewed towards university-educated males. Of additional note, they found that respondents' reported location did not match the location set by Facebook targeting much of the time.

The author of~\cite{freihardt2020can} uses Facebook and WhatsApp to survey inhabitants of Blantyre, Malawi on sanitation conditions.  The author finds difficulty reaching low-income households and speculates that it is because social media access generally requires a smartphone and an internet connection.

In~\cite{olamijuwon2021characterizing}, the author uses Facebook to recruit survey participants in Kenya, Nigeria, and South Africa. The author seeks to characterize the quality of the responses received by using Facebook as a method of recruitment. The author uses three ``attention check'' questions as a proxy measure of response quality and participant attention. The author finds about half of the participants pass all the attention checks. Additionally, the author gives evidence that passing attention checks correlates with age, sex, and other demographic characteristics. 

The authors of~\cite{rosenzweig2020survey} noticed evidence of link sharing  (snowball sampling) in their Kenya survey. However, they changed the survey link frequently to minimize the effect of link sharing minimized. As in our study, they offered a modest mobile airtime credit as compensation for completing their survey.

{Recently, in~\cite{apuke2024information}, researchers employed social media-driven chain referrals to examine misinformation and social media use in the context of public health campaigns in Nigeria.}

\subsection*{Article contributions}
{Our study examines the data from responses to remote surveys fielded by using ads via social media, a type of river sampling. Organically, word-of-mouth referrals, or snowball sampling, brought in a new group of respondents. We do not claim that the sample is representative. However, these two different online non-probability sampling processes have been subject to limited prior methodological investigation. Because we investigate an under-examined, hard-to-detect and measure, organic sampling process that is distinct from but may co-occur with river sampling, our findings are important for researchers who use social media to recruit survey participants. 

Due to the use of social media ads to obtain respondents, neither the river nor the snowball sampling process is researcher-controlled and the sampling frames are largely unknown. As a result, estimating the biases of each sample relative to the general adult population of the target geography and the means of mitigating the sampling biases face considerable challenges and are outside the scope of this introductory study of the emergent dual-sampling phenomenon.} 

The contributions of our article are: 
\begin{itemize}
    \item We provide evidence that the snowball sampling process resulted in a higher completion rate, a higher fraction of new respondents and a higher fraction of women respondents than the river sampling process.
    \item We show that the snowball sampling process resulted in shorter, quicker response behaviors for men.
    \item We give a heuristic for estimating the relative fraction of snowball sampling for a population containing both river and snowball sampling. 
    \item We make the Facebook and web survey data available for each of these six campaigns.
\end{itemize}

\section*{Methods}

\subsection*{Advertisement campaigns}

From late 2023 to early 2024, we ran six advertising campaigns targeting adults living in Zamfara, Nigeria to recruit participants to answer a survey about current and anticipated future food prices. We used a river sampling approach, leveraging the Meta advertising platform to deliver advertisements that linked to a survey hosted by LimeSurvey. Each ad campaign lasted five days. Table~\ref{tab:date} lists the dates that we ran the advertisements on Facebook and Instagram and collected responses on LimeSurvey. The advertisements consisted of one of eight different image-text pairs along with a link to the survey. A description of the images, the text used, and the cost of each one over the campaigns are in the appendix. We initially provided a payment of the equivalent of \$5 USD (roughly 4,000 Nigerian Naira, though conversion rates fluctuated significantly during the study period) in mobile air time to all respondents who completed the survey, and then in October we switched to a lottery-based incentive that paid the same amount to a randomly selected subset of 400 participants who completed a survey. We used the \textit{cost-per-click} pricing optimization scheme for the advertisements. Facebook users were able to take the survey multiple times, though no additional incentive was given for multiple responses from the same user during the same campaign.

\begin{table}[!ht]
\caption{{\bf Zamfara campaign dates.}}
\centering
 \begin{tabular}{|cccc|} 
 \hline
 Campaign&Start Advertising and Collecting &End Advertising&End Collecting\\ 
 \hline
Aug& Aug 25th 2023 & Aug 29th 2023 &  Sept 6th 2023  \\ 
Sept & Sept 21st 2023 & Sept 25th 2023  &  Oct 3rd 2023 \\ 
Oct & Oct 19th 2023 & Oct 23rd 2023 &  Oct 31st 2023 \\
Jan & Jan 27th 2024 & Jan 31st 2024 &  Feb 2nd 2024\\
Feb & Feb 13th 2024 & Feb 17th 2024 &  Feb 21st 2024\\
Mar & Mar 22nd 2024 & Mar 26th 2024 &  Mar 27th 2024\\
\hline
 \end{tabular}
 \label{tab:date}
\end{table}

\subsection*{Survey design}

We fielded the exact same survey across all advertisements and campaigns over time in order to control for any survey-specific response variation. The survey first collects demographic information including age, sex, state and district of current residence, religion, occupation, and income, among other features. Then, the survey asks respondents exactly what the current prices of rice, vegetables, and beef are in their local marketplaces, as well as what the respondents thought the prices would be six weeks from now, and why. The survey concludes with requests for administrative and contact information. The appendix contains the exact questions used.

\subsection*{Ethical considerations}

To obtain informed consent, we feature an informed consent statement with IRB and investigator contact information, research sponsor information, and a summary of the survey topics as well as the payment process. This was translated into Hausa, as was the survey itself, and the translation quality was verified via back translation. IRB approval was obtained via the Nigerian Federal Ministry of Health and the Health Media Lab in the United States. From the Nigerian Federal Ministry of Health, we received notice ``that the research described in the submitted protocol, consent form, advertisement and other participant information materials have been reviewed and given  expedited committee approval by the National Health Research Ethics Committee.'' From Health Media Lab, we are IRB \#929TSTI21. Ethical review and approval was also obtained via the Army Human Research Protections Office (HRPO). The reviews covered survey content, recruitment materials and approach, and incentive structures for all survey campaigns. 

\subsection*{Investigating sampling processes}
We identified that respondents arrived to our surveys through two different sampling mechanisms: some came to the survey through river sampling, by clicking on an ad on Facebook, and some came to the survey through snowball sampling, by clicking on an ad that has been re-posted or shared by a user.

In the course of this paper, we investigate if respondents from snowball sampling are statistically associated with different demographic distributions as well as participation and completion characteristics than those recruited via river sampling. In particular, we seek to examine the following potential characteristics of snowball sampling respondents:
\begin{itemize} 
    \item Higher completion rate
    \item Higher fraction of women respondents
    \item Higher fraction of respondents answering a survey for the first time
    \item Decrease in survey response length
    \item Decrease in time spent of survey
\end{itemize}

To compare river and snowball sampling, we divide the survey responses into two groups: a \textit{MIX} group that is the result of a combination of snowball and river sampling, and a \textit{SNOW} group that is entirely comprised of responses from participants drawn via snowball sampling.  We then attribute the differences between the \textit{MIX} and \textit{SNOW} groups to the respondents from river sampling in the \textit{MIX} group. Unfortunately, we cannot isolate a group that is entirely river sampled, because sharing behaviors were observed on Facebook during the initial campaigns.  

We construct the groups in the following manner: in August, September, October, January, and February, our web survey continued to receive responses days after the Facebook ad campaigns were finished running. Importantly, these were not completions of surveys begun previously, as we can identify by the start time (a data point that can be collected using LimeSurvey). Thus, these responses came from snowball sampling and these form the \textit{SNOW} group. The \textit{MIX} group consists of the responses to the survey while the Facebook ads were running, which were drawn from both river sampling \textit{and} snowball sampling.

In addition, during March, we ran two sets of ads; one set targeting men and the other targeting women using Meta's audience targeting option. Any women who completed a survey associated with the ads targeting men, as well as men who took a survey associated with the ads targeting women, form the \textit{SNOW} group. Men who took the ad targeting men and women who took the ad targeting women form the \textit{MIX} group. 

\subsection*{Analysis and measurement plan}
In this section, we define the quantities we compare between the \textit{MIX} and \textit{SNOW} groups and describe how we test for significance.

First, we define the completion rate as the number of full survey completions divided by the total number of completions of the initial informed consent. We model the completion rate as a beta distribution with $\alpha$ equal to the number of completed surveys and $\beta$ equal to the number of partially completed surveys. For beta distributions, we can calculate the p-values for significant tests exactly~\cite{bayesab}.

In our survey, we collected emails and phone numbers from survey participants. We categorize a respondent as ``returning'' if the provided email or phone number matches one that was given in a previous response during a previous campaign. A respondent is ``new'' if they provide an email or a number and neither one matches a previous email or phone number. Survey responses where the email or phone number matches an email or survey from the current campaign are labeled ``repeat'' responses. We model the fraction of the new respondents as a beta distribution with $\alpha$ equal to the number of new respondents and  $\beta$ equal to the number of returning respondents.

Next, we model the fraction of women respondents as a beta distribution with $\alpha$ equal to the number of women respondents and  $\beta$ equal to the number of men respondents.

The bulk of the survey asked questions about local food prices now and participants' predictions concerning food prices 6 weeks in the future—specifically, whether the prices of various goods would stay the same, increase, or decrease. Participants were also asked why they predicted these future price levels. We look at two metrics associated with the questions about food prices. First, the total time spent on answering the questions about the prices of food. Second, The number of total characters in the three free-response questions about the reasoning behind their six-week price predictions. In addition, we only look at the first time each respondent filled out a survey when comparing the time spent on and length of responses between the \textit{MIX} and \textit{SNOW} groups.

As we show in the results, the distribution of response time and length is skewed. As such, we consider the median as well as the mean when we test for significance. We perform these significance tests with 100,000 bootstrap sampling trials.  

Over the course of this article, we use a base significance level of 5e-2. For multiple tests, we use Bonferroni correction. For example, if we were doing a test across all six campaigns that we perform for men and women separately the Bonferroni corrected significance level would be 5e-2/12=4.17e-3. In each table we give the Bonferroni corrected significance level, labeled \textit{Bcsl}, and bold significant p-values. 

\subsection*{Estimating the proportion of respondents from snowball sampling in mixed responses}

Throughout our study, we examine a \textit{MIX} group that consists of participants recruited through both river sampling and snowball sampling. However, without additional tracking features or tools that may violate participant privacy, we are unable to identify whether a participant arrived at the survey by river or snowball sampling. Intuitively, if a \textit{MIX} group has a high fraction of snowball sampling, there will be minimal difference between the \textit{MIX} and corresponding \textit{SNOW} group. While we cannot determine which respondents came from snowball vs river sampling, we offer a heuristic to measure the relative fraction of snowball sampling in the \textit{MIX} group between the different campaigns. 

To do so, we examine the \textit{consent-to-click} ratio. This is the number of completed informed consents in the \textit{MIX} group divided by the number of clicks on ads reported by Meta. The intuition is that snowball sampling bypasses Meta's ad click count, but the users are counted in the number of consents. This does not identify which responses are from snowball sampling and which are not, but it enables us to estimate the extent of snowball sampling taking place. In the results, we present data in support of our intuition about the consent-to-click ratio from our February campaign.

\section*{Results}

\subsection*{Overview of results}

We list the cost to run each ad campaign via Meta (cost), the clicks Meta reported (clicks), the number of total surveys started (consents), and the number of completed surveys (completions) from each of the six campaigns in Table~\ref{tab:overview}. 

\begin{table}[!ht]
\caption{{\bf Overview statistics of the campaigns.}}
\centering
 \begin{tabular}{|ccccc|} 
 \hline
 Campaign& Cost&Clicks & Consents& Completions  \\ 
 \hline
Aug &1796.58 & 16225& 1290 & 886 \\ 
Sept&1794.58 & 13715  & 1992 & 1610\\ 
Oct &349.05 & 8792&2177 & 1875 \\
Jan &46.62 & 2274 & 1654 & 1521\\
Feb &371.76 & 11326 & 1413 & 1210\\
Mar & 275.04 & 10115 & 880 & 699 \\
\hline
 \end{tabular}
 \label{tab:overview}
\end{table}

Additionally, Table~\ref{tab:mar_overview} contains the statistics for the ads that targeted men and the ads that targeted women in March. Ads targeting men have significantly more clicks, consents, and completions for the same cost.  

\begin{table}[!ht]
\caption{{\bf Number of men and women completions of the surveys linked to the targeted ads.}}
\centering
 \begin{tabular}{|ccccc|}
  \hline
    Target sex&Cost&Clicks& Consents & Completions  \\ 
   \hline
    men&137.52&7161&785&640\\
       women &137.52&2954& 95& 59\\
   \hline
  \end{tabular}
  \label{tab:mar_overview}
\end{table}

\subsection*{Effect of snowball sampling on completion rate}

We list the completed consents and survey completions that occurred while the Facebook ads were running (\textit{MIX}) and when the Facebook ads were not running (\textit{SNOW}) in Table~\ref{tab:compl}. In August, September, and February, a majority of the completions occur while the ads are running, and then the number of completions tappers off after the ads are completed, as expected. 

However, this is not the case in October and January. We believe these two instances are the results of different events and mechanisms. In the case of October, we facilitated incentive payment to the final tranche of September respondents on October 24th, 2024. Starting that day, our campaign's social media page saw a surge of attention and we received a new wave of completed surveys. Therefore, paying September respondents (who participated in a much larger and more heavily funded ad campaign than in October) likely influenced the number of participants who shared October survey links after encountering them via advertisements. With respect to the January campaign, Meta only ran our ads for four hours a day and only used a small portion of the allocated budget. As a result, we had substantially less recruitment via river sampling in general during January. At the same time, because of a brand recognition and reputation that had been built up over time, our January campaign ads were widely shared by audience members who encountered them. This suggests a cumulative effect related to community awareness of the survey campaigns on the incidence and amount of snowball sampling resulting from river sampling.

\begin{table}[!ht]
\caption{{\bf Consents and completions (compls) for the \textit{MIX} and \textit{SNOW} groups.}}
\centering
 \begin{tabular}{|ccccc|} 
 \hline
 Campaign & \textit{MIX} consents & \textit{MIX} compls & \textit{SNOW} consents & \textit{SNOW} compls\\ 
 \hline
Aug & 1220 & 822&70& 64\\ 
Sept & 1804&1449&188&161\\
Oct & 504 & 398 & 1673&1477 \\
Jan  &660 & 599& 994&922 \\ 
Feb  &1341 & 1144& 72&66 \\ 
  \hline
 \end{tabular}
 \label{tab:compl}
\end{table}

We test the hypothesis that the completion rate of \textit{MIX} is equal to that of \textit{SNOW} with the alternative hypothesis that the \textit{MIX} completion rate is lower than the \textit{SNOW} completion rate. Table~\ref{tab:compl_rate} lists the completion rates of the \textit{MIX} and \textit{SNOW} groups as well as the p-values of associated significance tests. 

\begin{table}[!ht]
\caption{{\bf The \textit{MIX} and \textit{SNOW} groups' completion rate (compl rate) and the p-values with alternative hypothesis that \textit{MIX}'s compl rate $<$ \textit{SNOW}'s compl rate.} Bcsl:1e-2, Significant p-values in bold.}
\centering
 \begin{tabular}{|c|c|c|c|} 
    \hline
 Campaign & \textit{MIX} compl Rate  & \textit{SNOW} compl Rate & P-value \\ 
 \hline
Aug & 0.674& 0.914& \textbf{$<$1e-5} \\ 
Sept & 0.803&0.856 & 3.14e-2 \\
Oct & 0.790& 0.883 & \textbf{$<$1e-5} \\
Jan  & 0.908&0.928& 7.37e-2 \\ 
Feb & 0.853&0.917 & 4.42e-2 \\ 
 \hline
 \end{tabular}
 \label{tab:compl_rate}
\end{table}

In all months, the completion rate of \textit{MIX} is lower than that of \textit{SNOW} and in August and October, the completion rate of \textit{MIX} is significantly lower than that of \textit{SNOW}. Thus, there is evidence that snowball sampling leads to a higher completion rate.

\subsection*{New, returning, and repeat respondents}

In Table~\ref{tab:return}, we list the number of new respondents, returning respondents (respondents who filled out a survey in a previous campaign), repeat responses (surveys completed by participants who already filled one out during the campaign), and surveys with no valid phone number or email address (unknown). One observation is the fraction of returning respondents grows with each successive campaign. In fact, during March there are more returning respondents than new respondents. This indicates that we are reaching a saturation point of audience members from the study geography willing to continue participating in surveys as currently advertised, structured, and incentivized.

\begin{table}[!ht]
\caption{{\bf Number of surveys from new and returning respondents as well as number of repeat surveys and surveys without respondent information.}}
\centering
 \begin{tabular}{|ccccc|} 
 \hline
 Campaign & New & Returning & Repeat & Unknown \\ 
 \hline
Aug & 791&-&73&22 \\ 
Sept & 1130& 228&210&42\\
Oct & 1340&384&121&30  \\
Jan  &923 & 437& 141& 20 \\ 
  Feb  &631 & 462& 106& 11 \\ 
  Mar  &215&388&86&10 \\ 
   \hline
 \end{tabular}
 \label{tab:return}
\end{table}

\subsection*{Effect of snowball sampling on reaching women and new respondents}

We begin by showing the number of respondents in the \textit{MIX} and \textit{SNOW} groups broken up by self-reported sex and new vs returning in Table~\ref{tab:ret_sex_MIX}. Unsurprisingly, far more men took the survey than women. This is consistent with the demographics of Meta platform users in Zamfara state, which are viewable via Meta's ad management system audience analysis tools and which skew dramatically toward men.

\begin{table}[!ht]
\caption{{\bf Number of respondents broken up by new vs returning, sex, and \textit{MIX} vs \textit{SNOW} group.}}
\centering
 \begin{tabular}{|ccccccccc|} 
 \hline
 &\multicolumn{4}{c}{New}&\multicolumn{4}{c|}{Returning}\\
  &\multicolumn{2}{c}{Men}&\multicolumn{2}{c}{Women}&\multicolumn{2}{c}{Men}&\multicolumn{2}{c|}{Women}\\
 Campaign & \textit{MIX} & \textit{SNOW} & \textit{MIX} & \textit{SNOW}  & \textit{MIX} & \textit{SNOW}& \textit{MIX} & \textit{SNOW} \\ 
 \hline
Aug & 664 & 50& 65&12&- & -& -& - \\ 
Sept & 762&86&253&29&173&24&25&6\\
Oct & 99 & 824 & 22& 395&183&101&55&45  \\
Jan  &197 & 470& 74& 182&209&130&37&61 \\ 
Feb  &408 & 28& 186& 9&358&15&83&6 \\ 
 \hline
 \end{tabular}
 \label{tab:ret_sex_MIX}
\end{table}

First, we look at the proportions of women taking the surveys. In Table~\ref{tab:pvalue_women}, we list the fraction of women in the \textit{MIX} and \textit{SNOW} groups broken up by new and returning respondents as well as the p-value that the fraction of women in the \textit{MIX} and \textit{SNOW} groups are equal with the alternative hypothesis that proportion of women in the \textit{SNOW} group is higher than in the \textit{MIX} group. In October, new respondents in the \textit{SNOW} group had a significantly higher proportion of women respondents than new respondents in the \textit{MIX} group. In January, returning respondents in the \textit{SNOW} group had a significantly higher fraction of women respondents than among returning respondents in the \textit{MIX} group. Also of note is that the fraction of women is higher in the \textit{SNOW} group than in the \textit{MIX} group in all cases except for new respondents in February. Thus, there is evidence that snowball sampling consistently leads to a higher fraction of women respondents. 

\begin{table}[!ht]
\caption{{\bf Snowball sampling effect on finding women.} Bcsl: 5.56e-3, significant p-values in bold.}
\centering
 \begin{tabular}{|ccccccc|} 
 \hline
  & \multicolumn{3}{c}{New} & \multicolumn{3}{c|}{Returning} \\
 Campaign & \textit{MIX} frac  & \textit{SNOW} frac & P-value &\textit{MIX} frac  & \textit{SNOW} frac & P-value \\
  \hline
    Aug & 0.089 & 0.194& 9.49e-3&-&-&-\\ 
Sept & 0.249 & 0.252& 0.485&0.126& 0.200&0.165\\ 
Oct & 0.182 & 0.324 & \textbf{3.3e-4} & 0.231 & 0.308 & 4.86e-1\\
Jan  &0.273 &0.279  & 0.421& 0.150&0.319 &  \textbf{1e-5} \\
  Feb  &0.313 &0.243  & 0.833& 0.188&0.286 & 0.164 \\
  \hline
 \end{tabular}
 \label{tab:pvalue_women}
\end{table}

Next, we analyze the fraction of new respondents. Table~\ref{tab:pvalue_new} enumerates the fraction of new respondents in the \textit{MIX} and \textit{SNOW} groups broken up by self-reported sex as well as the p-value that the fraction of new respondents in the \textit{MIX} and \textit{SNOW} groups are equal with the alternative hypothesis that the fraction of new respondents in the \textit{SNOW} group is higher than the \textit{MIX} group. In October, there are significantly more new men and women respondents in the \textit{SNOW} group, and in January there are significantly more new men respondents in the \textit{SNOW} group. The fraction of new respondents is also higher in the \textit{SNOW} group for women in January and for men in February, but not significantly so. On the other hand, the \textit{MIX} group for men and women in September contains more new respondents and the \textit{MIX} women group in February contains more respondents, however not to significant levels. Demonstrably, there is evidence that snowball sampling leads to a higher fraction of new respondents.  

\begin{table}[!ht]
\caption{{\bf Snowball sampling effect on finding new people.} Bcsl: 4.16e-3, significant p-values in bold.}
\label{tab:pvalue_new}
\centering
 \begin{tabular}{|ccccccc|} 
 \hline
  & \multicolumn{3}{c}{Men} & \multicolumn{3}{c|}{Women} \\
 Campaign & \textit{MIX} frac  & \textit{SNOW} frac & P-value &\textit{MIX} frac  & \textit{SNOW} frac & P-value \\ 
 \hline
Sept & 0.815 & 0.782& 0.785& 0.910 & 0.829 & 0.908 \\ 
Oct & 0.351 & 0.891 & \textbf{$<$1e-5} & 0.286 & 0.898 & \textbf{$<$1e-5} \\
Jan  &0.485 &0.783  & \textbf{$<$1e-5}&0.667 &0.749  & 5.67e-2 \\
Feb  &0.533 &0.651  & 6.00e-2 &0.691 &0.600  & 0.754 \\
\hline
 \end{tabular}
\end{table}

In Table~\ref{tab:target_first_return}, we list the number of new and returning men and women who completed the surveys linked to the ads targeting men and those who completed surveys linked to the ads targeting women. As previously mentioned, we used Meta's audience targeting capabilities to specifically target men and women social media users with distinct ads and survey links, and examined the sets of completions for each survey. Importantly, we performed this experiment in March 2024, \textit{after} our campaigns had begun to exhibit a pattern of significant completions from snowball sampling (i.e., after a cumulative brand ``reputation'' may have been established in Zamfara). In examining the survey responses, we observe that, perhaps surprisingly, many more women complete surveys from the ads targeting men. This means that at least 80 of 109 (over 70\%!) of women came to our ad from snowball sampling in March.  

\begin{table}[!ht]
\caption{{\bf Number of men and women completions of the surveys linked to the targeted ads.}}
\centering
 \begin{tabular}{|ccccc|}
  \hline
  & \multicolumn{2}{c}{New} & \multicolumn{2}{c|}{Returning}\\

    Target sex & Men & Women & Men & Women  \\ 
   \hline
    Men&149&33&323&47\\
    Women & 14&19& 8&10\\
   \hline
  \end{tabular}
  \label{tab:target_first_return}
\end{table}

\subsection*{Effect of snowball sampling on response speed and length}

We are unable to directly measure response quality via attention check or quality control questions, as response quality was not a priority during earlier phases of data collection, and we later opted not to change survey instruments to add new questions so as to be able to compare completion and attrition rates from different campaigns. However, below we investigate if respondents in the \textit{SNOW} group spend significantly less time or have significantly shorter responses than the \textit{MIX} group. We only consider respondents who are answering the survey for the first time. Additionally, we examine these factors broken out by sex.

\subsubsection*{Men}

We show the response time and length for men in the \textit{MIX} and \textit{SNOW} groups for the month of August in Fig~\ref{fig:aug_men}. Plots for all other months are given in the appendix. In Table~\ref{tab:men_MIX} and \ref{tab:men_SNOW}, we tabulate the mean and median time and response length for men in the \textit{MIX} group and men in the \textit{SNOW} group, respectively. With a small number of exceptions (specifically, response time in September and response length in January), on average the men in the \textit{SNOW} group spent less time on and gave shorter responses to the food price questions than those in the \textit{MIX} group. 

\begin{figure}[!ht]
\includegraphics[width=\textwidth]{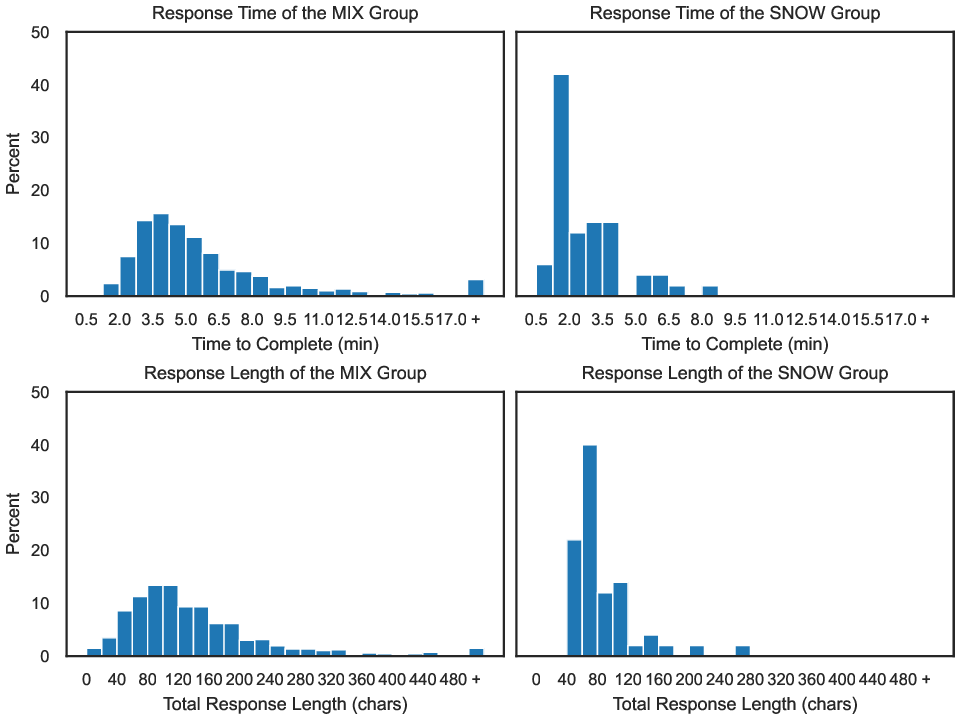}
\caption{{\bf Response time and length for men in the \textit{MIX} and \textit{SNOW} group in August}. The x-axis label `$+$' refers to responses over 17 min for response time and over 480 characters for response length.} 
\label{fig:aug_men}
\end{figure}

\begin{table}[!ht]
\caption{{\bf Average completion time and response length for questions on food price for men respondents in the \textit{MIX} group.}}
\centering
 \begin{tabular}{|cccccc|} 
 \hline
 Campaign & Count &  Mean time&Median time& Mean length&Median length\\ 
 \hline
Aug  & 664 & 6.03 & 4.81 & 148.9 & 115.0 \\ 
Sept & 762 & 4.93 & 3.98 & 123.2 & 100.0 \\ 
Oct & 99 & 6.48 & 5.17 & 131.8 & 103.0 \\
Jan  & 197 & 4.06 & 3.04 & 126.6 & 96.0 \\
Feb  & 408 & 4.50 & 3.41 & 117.5 & 84.0 \\
Mar  & 149 & 5.06 & 3.80 & 125.4 & 91.0 \\
\hline
 \end{tabular}
 \label{tab:men_MIX}
\end{table}

\begin{table}[!ht]
\caption{{\bf Average completion time and response length for questions on food price for men respondents in the \textit{SNOW} group.}}
\centering
 \begin{tabular}{|cccccc|} 
 \hline
 Campaign & Count & Mean time&Median time& Mean length&Median length\\ 
 \hline
Aug  & 50 & 2.80 & 2.06 & 86.9 & 74.0 \\ 
Sept & 86 & 6.00 & 4.38 & 123.0 & 92.5 \\ 
Oct & 824 & 4.44 & 3.55 & 95.2 & 79.0 \\
Jan  & 470 & 3.24 & 2.30 & 131.6 & 102.0 \\
Feb  & 28 & 3.23 & 2.78 & 96.3 & 70.5 \\
Mar  & 14 & 4.91 & 3.75 & 49.5 & 41.0 \\
\hline
 \end{tabular}
 \label{tab:men_SNOW}
\end{table}

In Table~\ref{tab:men_pvaule}, we give the p-value of the hypothesis test that the mean/median of the times or response lengths are equal with the alternative hypothesis that the men in the \textit{MIX} group take longer or write longer answers. In August and October, the men in the \textit{MIX} group take significantly longer and write significantly longer answers. In January, the mean time spent on the questions was significantly higher for men in the \textit{SNOW} group.  In March, the men in the \textit{MIX} group wrote significantly longer answers. Thus, there is \textbf{strong} evidence that snowball sampling leads to shorter and faster responses from men.

\begin{table}[hp]
\caption{{\bf P-values that the average completion time and response length are equal for men between the \textit{MIX} and the \textit{SNOW} group with the alternative hypothesis that \textit{MIX}$>$\textit{SNOW}.} Bcsl: 4.16e-3, significant p-values in bold.}
\centering
 \begin{tabular}{|ccccc|} 
 \hline
 campaign  &  Mean time&Median time& Mean length&Median length\\ 
 \hline
Aug  & \textbf{1e-5}& \textbf{1e-5} & \textbf{1.4e-4} & \textbf{1e-5} \\ 
Sept & 0.982 & 0.846 & 0.511 & 0.184 \\ 
Oct & \textbf{4e-5} & \textbf{3e-5} & \textbf{4e-5} & \textbf{1.58e-3} \\
Jan  & 7.81e-3 & \textbf{1.31e-3} & 0.743 & 0.834 \\
Feb  & 2.33e-2 & 0.157 & 0.192 & 0.170 \\
Mar  & 0.482 & 0.440 & \textbf{2.9e-4} & \textbf{2.19e-3} \\
\hline
 \end{tabular}
 \label{tab:men_pvaule}
\end{table}

\subsubsection*{Women}

We show the response time and length for women in the \textit{MIX} and \textit{SNOW} groups for the month of January in Fig~\ref{fig:jan_women}. Plots for all the months are given in the appendix.

\begin{figure}[!ht]
\includegraphics[width=\textwidth]{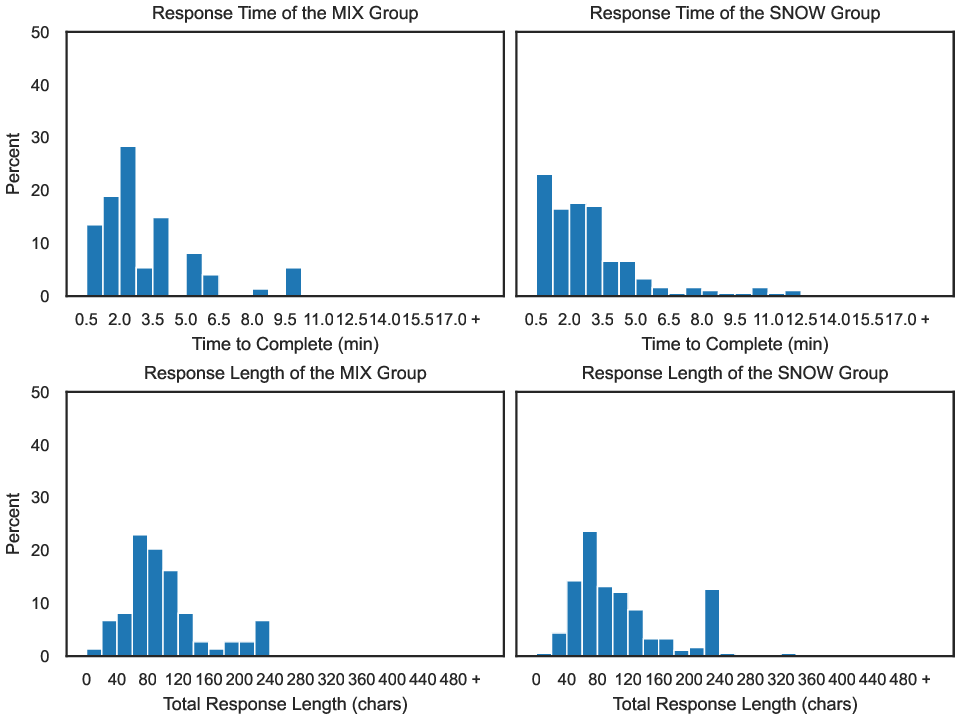}
\caption{{\bf Response time and length for women in the \textit{MIX} and \textit{SNOW} group in January}. The x-axis label `$+$' refers to responses over 17 min for response time and over 480 characters for response length.}
\label{fig:jan_women}
\end{figure}

In Table~\ref{tab:women_MIX} and \ref{tab:women_SNOW} we tabulate the mean and median time and response length for women in the \textit{MIX} group and women in the \textit{SNOW} group respectively. Unlike for men, the \textit{MIX} group's average response time/length is only greater approximately 54\% of the time.

\begin{table}[hp]
\caption{{\bf Average completion time and response length for questions on food price for women respondents in the \textit{MIX} group.}}
\centering
 \begin{tabular}{|cccccc|} 
 \hline
 Campaign & Count &  Mean time&Median time& Mean length&Median length\\ 
 \hline
Aug  & 65 & 5.21 & 3.94 & 125.8 & 105.0 \\ 
Sept & 253 & 3.66 & 2.87 & 100.4 & 96.0 \\ 
Oct & 22 & 5.53 & 4.92 & 120.2 & 92.0 \\
Jan  & 74 & 3.09 & 2.29 & 100.4 & 87.5 \\
Feb  & 186 & 3.24 & 2.20 & 95.3 & 75.0 \\
Mar  & 19 & 4.04 & 3.19 & 92.2 & 72.0 \\
\hline
 \end{tabular}
 \label{tab:women_MIX}
\end{table}

\begin{table}[hp]
\caption{{\bf Average completion time and response length for questions on food price for women respondents in the \textit{SNOW} group.}}
\centering
 \begin{tabular}{|cccccc|} 
 \hline
 Campaign & Count & Mean time&Median time& Mean length&Median length\\ 
 \hline
Aug  & 12 & 3.02 &1.62 & 112.3 & 70.5 \\ 
Sept & 29 & 3.64 & 3.29 & 95.7 & 95.0 \\ 
Oct & 395 & 4.09 & 3.29 & 96.9 & 84.0 \\
Jan  & 182 & 2.98 & 2.41 & 110.7 & 87.5 \\
Feb  & 9 & 3.62 & 2.62 & 114.2 & 126.0 \\
Mar  & 149 & 3.82 & 3.49 & 134.0 & 108.0 \\
\hline
 \end{tabular}
\label{tab:women_SNOW}
\end{table}

In Table~\ref{tab:men_pvaule}, we give the p-value of the hypothesis test for equal mean/median response times/lengths with the alternative hypothesis that the women in the \textit{MIX} group take more time or compose longer responses. Only in August is there a significant difference and only for the median response time. Thus, there is \textbf{weak, if any,} evidence that snowball sampling leads to faster or shorter responses among women.

\begin{table}[hp]
\caption{{\bf P-values that the average completion time and response length are equal for women between the \textit{MIX} and the \textit{SNOW} group with the alternative hypothesis that \textit{MIX}$>$\textit{SNOW}.} Bcsl: 4.16e-3, significant p-values in bold.}
\centering
 \begin{tabular}{|c|c|c|c|c|c|} 
 \hline
 Campaign  &  Mean time&Median time& Mean length&Median length\\ 
 \hline
Aug  & 1.16e-2& \textbf{5.3e-4} & 0.338 & 1.65e-2 \\ 
Sept & 0.518 & 0.871 & 0.341 & 0.465 \\ 
Oct & 3.98e-2 & 8.20e-3 & 7.32e-2 & 0.311 \\
Jan  & 0.358 & 0.697 & 0.888 & 0.521 \\
Feb  & 0.713 & 0.748 & 0.852 & 0.981 \\
Mar  & 0.353 & 0.745 & 0.983 & 0.948 \\
\hline
 \end{tabular}
 \label{tab:women_pvaule}
\end{table}

\subsection*{Consent-to-click ratios}

In February, we ran four copies of the same ad. Table~\ref{tab:feb_cc} contains the clicks, consents in the \textit{MIX} group, the consent-to-click ratios, and the consents from the \textit{SNOW} group from the four ads. While the clicks are within a reasonable variation of each other, Harvest 4 has much higher consents than the other three ads in the \textit{MIX} and in the \textit{SNOW} group. This suggests that the survey associated with the Harvest 4 ad saw a greater number of respondents recruited via snowball sampling, and subsequently supports that the consent-to-click ratio is correlated to the fraction of snowball sampling. It also lends credence to the framing of snowball sampling as a random process arising from river sampling with some latent probability.

\begin{table}[!ht]
\caption{{\bf February ad cost, clicks, \textit{MIX} and \textit{SNOW} consents, and consent-to-click ratio (CC Ratio).}}
\centering
 \begin{tabular}{|cccccc|} 
 \hline
 Ad&Cost&Clicks & \textit{MIX} Consents& CC Ratio& \textit{SNOW} Consents \\ 
 \hline
Harvest 1&92.94 & 2826 & 288 &0.102 & 10\\ 
Harvest 2&92.94 & 2720 & 327 &0.120 & 4\\ 
Harvest 3&92.94 & 2877 & 279 &0.097 & 5\\
Harvest 4&92.94 & 2903 & 447 &0.154 & 53 \\
  \hline
 \end{tabular}
 \label{tab:feb_cc}
\end{table}

In Table~\ref{tab:cc}, we tabulate the consent-to-click ratios for the August, September, October, January, and February campaigns. The January campaign has a exceptionally high ratio. August and October have relatively low ratios.

\begin{table}[!ht]
\caption{{\bf Consents and clicks for the \textit{MIX} group August through February.}}
\centering
 \begin{tabular}{|cccc|} 
 \hline
 Campaign&Clicks & Consents& CC Ratio  \\ 
 \hline
Aug  & 16225& 1220 & 0.075 \\ 
Sept & 13715  & 1804 & 0.132\\ 
Oct & 8792&504& 0.057 \\
Jan & 2274 & 660 & 0.290\\
Feb & 11326 & 1341 & 0.118\\
\hline
 \end{tabular}
 \label{tab:cc}
\end{table}
Finally, we give the consent-to-click ratios of the March ads targeting men and the ads targeting women in Table~\ref{tab:mar_cc}. We see that the ads targeting women have a very low ratio. From Table~\ref{tab:target_first_return} we have that 22 of the completions were men which means that at least 23\% of consents were men. Possible reasons for a low consent-to-click ratio and a high number of known snowball sampling are \textit{MIX} groups have a lot of snowball sampling, Meta's targeting does not match self-reported gender a portion of the time, or women are more likely to click an ad and not consent than men. Regardless of the reason, the consent-to-click ratio when using targeting may not be directly comparable and merits future examination.

\begin{table}[!ht]
\caption{{\bf Consents and clicks for the targeted ads from March while the Facebook ads were running.}}
\centering
 \begin{tabular}{|cccc|}
  \hline
    Target sex&Clicks & Consents& CC Ratio  \\ 
   \hline
    Men&7161&780&0.109\\
    Women &2954& 95& 0.032\\
   \hline
  \end{tabular}
  \label{tab:mar_cc}
\end{table}

\section*{Discussion}

We observed an emergent secondary sampling mechanism intermixed with our primary river sampling approach, that appears to exhibit several statistically significant differences in responses, response behaviors, and sample composition. We measured the difference between the \textit{MIX} and the \textit{SNOW} group as a proxy for comparing the differences between river sampling and snowball sampling. In Table~\ref{tab:cc}, we see that August and October have the lowest consent-to-click ratio and, we conjecture, the lowest fraction of snowball sampling in the \textit{MIX} group. As a result, the signal between river and snowball sampling is the strongest in those months. As a result, only August and October had significant differences between the \textit{MIX} and \textit{SNOW} groups for completion rate and the only months to have significantly mean and median lower time spent and shorter answers on the food questions for men. Meanwhile, the \textit{MIX} groups in the other campaigns have a high enough fraction of users from snowball sampling that it is difficult to detect differences between the \textit{MIX} and \textit{SNOW} groups. 

In summary, we present evidence that the users from snowball sampling have a higher completion rate, that snowball sampling leads to a higher fraction of women and new respondents, that open-text responses from men obtained by snowball sampling are shorter in length, and that survey responses were completed more quickly than comparable responses from men obtained by river sampling.

A possible reason that snowball sampling is better at obtaining women and new respondents is that users, and women especially, are more willing to click a post from an acquaintance than an ad. 

As for the decrease in response time and length, it is not unreasonable that there is a correlation between users willing to click on a survey from a stranger and those willing to spend time answering the questions.

When using gender-specific targeting, our data suggests that notably more completions by women arise from snowball sampling than from river sampling. Additionally, a higher fraction of snowball sampling in the women \textit{MIX} group would obscure the signal between river and snowball sampling. This is a possible reason why we could not detect a significant difference in length and speed of the responses between the women \textit{MIX} and \textit{SNOW} groups. This finding is consistent with past research on the use of respondent referrals, including respondent-driven sampling (RDS), to reach hard-to-reach sub-populations more effectively and efficiently~\cite{platt2006rds}. 

As a final observation, in February, we ran the same ad four times and got significantly different results. Without a better way to disentangle the river sampling and snowball sampling AB testing, or determining the most effective ad, is difficult. From the February data, we would conclude that Harvest 4 is the best ad of the four but, in reality, all the ads are the same so there is no meaningful difference. Thus, the boost from snowball sampling can obscure the signal of higher completions from a better ad. 

\section*{Conclusion and future work}

{
Surveying via social media presents challenges to current surveying methods. As one example, the secondary sampling we have observed may not be a suitable mechanism to adapt to a respondent driven-sampling (RDS) use case for two reasons. First, the sampling frame is not known. RDS relies on a sampling mechanism that enables access to the entirety of the population under study, to build an approximate probability sample of that population. For instance, seeds may be selected using time-location sampling (TLS) of a hidden population that congregates in specific locales at specific times, with in-person coupons used to facilitate chain referrals, under the assumption that all members of the target population are in the selected geography and have in-person contact with one another. In the case of our study, the subset of individuals in a target geography that is online and active on social media is not precisely known. We may only approximate this sub-population and, in doing so, would risk introducing new sampling biases. 

Second, RDS uses a specific number of referrals per respondent to grow a sample and relies on this limit to maintain sample balance and facilitate network propagation. These successive samples are easily tracked and moderated in most RDS use cases via coupon codes that identify the referrer-respondent (by code if not by name), and subsequently associate the new respondent with the referrer from the previous wave. This tracking capability also enables model calibration in cases where referrals may be imbalanced. However, the secondary sampling mechanism examined in our study may be arbitrarily one-to-many, facilitated via a post on a social media profile, a message to a group via WhatsApp, or another messaging channel.

Future studies can mitigate both of these challenges by: (1) collecting detailed data on the demographics of social media users in the target geography, for example via third-party advertisers; (2) using a digital tracking mechanism to follow referral chains emerging from initial survey participants. We caution that both of these mitigation strategies risk violations of participant privacy and/or research ethics protocols, and encourage researchers to collaborate closely with their IRBs as well as consult the relevant privacy laws within study jurisdictions.}

This paper contributes to non-probability sampling research and to digital survey research in Africa by examining the emergence of mixed sampling processes from initial river samples when performing social media-based survey recruitment in Nigeria. Previously, researchers using river sampling have endeavored to restrict snowball sampling from taking place, owing to the opacity and lack of control over the process. On the other hand, we believe this phenomenon offers both advantages and disadvantages, and may present new opportunities for non-probability sampling research. 

Snowball sampling has the benefit of efficiently providing the researcher with a larger pool of respondents. Indeed, in our study, we find that this secondary sampling process was a recurrent phenomenon, and dramatically increased the number of respondents to our surveys, including new respondents and those from hard-to-reach populations in the study geography. It also dramatically improved our overall cost-per-completion. On the other hand, the distribution of survey response length and speed is different between river and snowball sampling, and between male and female respondents under snowball sampling. The snowball sampling process is also relatively opaque and was left generally unrestricted in our experiments, as we did not seek to counter any specific sample biases or implement additional controls. 

Researchers seeking to build large samples efficiently or seeking to access hard-to-reach populations more effectively may find success leveraging a hybrid sampling approach enabled by incentivized digital surveys fielded using online advertisements. For example, researchers could leverage ad imagery and text that cater to a particular sub-population, and then explore whether snowball sampling processes that emerge over multiple study waves remain largely within or among group members. 

However, they should do so in a manner that does not jeopardize the validity of the study design or findings, nor risk incentivizing biased or dishonest responses. Mitigation measures could include attention checks and additional established mechanisms to validate respondents' sincerity as well as their self-reported membership in specific groups. From a design perspective, as long as the study design and topic do not induce participant risk via concealment, researchers may find it is best to not publicly specify a subgroup of interest to encourage honest participation and to reward all participants even if they are not members of a specific group.

We also believe that researchers should be cautious and thoughtful when implementing a design that, if not carefully implemented, may raise tricky ethical concerns. Although surveys are generally considered a lower-risk form of human subjects research, sensitive survey topics can still be risky to promote, and marginalized groups are particularly vulnerable. Users may inadvertently reveal sensitive or vulnerable information about themselves by sharing a survey whose published findings are sensitive or controversial. For this reason, the potential for public sharing and snowball sampling should also be communicated to ethical review boards when designing studies that use river sampling, and in some cases, additional risk mitigation measures may be recommended.

Given these observations, we note that past literature~\cite{rosenzweig2020survey} has identified basic methods to control digital snowball sampling referrals, and that leveraging these methods while deliberately permitting \textit{some} referrals may enable researchers to tailor the mixture of participants from river sampling and snowball sampling to improve the composition of the overall sample (for instance, under a quota sampling regime, or in other cases that include specific demographic or other targets). 

We also note that for the statistics in this paper, we did not have a group of respondents drawn \textit{solely} from river sampling. However, a pure river sampling group would have a much clearer signal. Future work could seek to estimate which respondents are from snowball sampling. Data and code available upon reasonable request.

\section*{Acknowledgments}
We thank Abdullah Rahmanyar for his expertise in implementing the social media campaigns used to engage with the target audiences.

The material is based upon work supported by Army Contracting Command, DARPA, and ARO under Contract No. W911NF-21-C-0007.

Any opinions, findings and conclusions or recommendations expressed in this material are those of the authors and do not necessarily reflect the views of the Army Contracting Command, DARPA, and ARO. 

\section*{Distribution Statement}
Distribution Statement ``A'' (approved for public release: distribution unlimited).

\section*{Appendix}

\subsection*{Ads fielded during the campaigns.}
\label{S1_Appendix}
We list the ads along with their text and a description of the accompanying image in Table \ref{tab:ads}. We cannot show the images because they are copyrighted. We obtained permission to use the images for our research. We list how much money we spent to advertise each ad through Facebook advertisements in Table \ref{tab:campaign_money} during each campaign. We would set a budget for each ad and Meta would spend a certain amount of the money (usually less but occasionally more) on the ad. For example, we set a budget of \$125 for each ad in October and Meta only fielded three of the ads, taking ~\$116 for ads Store, Harvest, and Transport.
\begin{table}[ht!]
\caption{{\bf Ad text and images}}

\centering
 \begin{tabular}{|c|p{6cm}|p{6cm}|} 
 \hline
 Ad&Text&Image description\\ 
 \hline
 Food&Delicious! What is the cost of basic food items at your local market? Earn 5000 NGN of mobile airtime by taking a survey!& Baskets and bags of food.\\ 
 \hline
 Fuel&Fuel stop! How will transportation issues impact food markets? Earn 5000 NGN of mobile airtime by taking a survey!& Two people on motorcycles at a gas station.\\ 
  \hline
 Store&Sold out! What do future prices look like for groceries in your area? Earn 5000 NGN of mobile airtime by taking a survey!& A man walking through a grocery store with empty shelves. \\
  \hline
 Fence&Secure! Will security affect prices of food at your local marketplace? Earn 5000 NGN of mobile airtime by taking a survey!& A barbed wire fence in front of a field. \\
  \hline
 Market&Local market! What is it like to shop for food at local marketplaces? Earn 5000 NGN of mobile airtime by taking a survey! & A woman in front of a food stand looking at the camera. \\
 \hline 
 Harvest&Harvest time! How does the crop yield look this season? Earn 5000 NGN of mobile airtime by taking a survey! & Several Women in a field harvesting food. \\
 \hline
 Transport&Farm to market! What is it like to shop for food at local marketplaces? Earn 5000 NGN of mobile airtime by taking a survey! & A busy marketplace. Prominent is a person on a motorcycle with a basket of goods on their head. \\
 \hline
 Mud& Stuck in the mud! How does the season change food prices? Earn 5000 NGN of mobile airtime by taking a survey! & Several people trying to get a car unstuck from mud that goes halfway up its wheels.  \\
 \hline
 \end{tabular}
 \label{tab:ads}
\end{table}
The ad text in Hausa is available upon reasonable request.
\begin{table}[ht!]
\caption{{\bf Money spent on each ad for each campaign}}

\centering
 \begin{tabular}{|c|c|c|c|c|c|c|c|c|} 
 \hline
 Campaign/Ad & Food&Fuel&Store&Fence&Market&Harvest&Transport&Mud\\ 
 \hline
Aug  &  224.57&  224.58&  224.57 &  224.57 &224.57 & 224.57& 224.57 & 224.58 \\ 
 \hline
Sept  &  112.15&  112.16& 112.15 &  665.37 & 243.71 & 324.73& 112.16 & 112.15\\ 
\hline
Oct   &  0&  0&  116.36 &  0 &0 & 116.36& 116.33 & 0 \\
  \hline
Jan  &   0&  23.35 &  0 &  0 &0 & 23.37 & 0 & 0\\
  \hline
Feb  &  0&  0&  0 &  0 &0 & 371.76& 0 & 0\\
\hline
Mar  &  0&  45.84&  0 &  45.84 & 0 & 91.68& 0 & 91.68\\
\hline
 \end{tabular}
 \label{tab:campaign_money}
\end{table}

\subsection*{The survey provided to the respondents}
\label{S2_Appendix}
We enumerate the questions from the survey. The survey was given in Hausa. This is the translation of the survey in English:
\begin{enumerate}
    \item We would like to ask you some questions. To show our appreciation we will randomly select 400 participants to receive \$5 (USD) worth of mobile data. \{Research participant information and consent form.\} Do you want to continue?
    \item What is your age? Please provide your age in years; example 25.
    \item What is your gender?
    \item What is the name of the village, town, or city in which you live?
    \item Is your residence located in an area that is urban or rural?
    \item What is your religious background?
    \item How many children do you have? If you have no children, answer 0 (zero).
    \item What is your occupation?
    \item  What is the price of 1 cup of local rice (sold loose) in your local market?
    \item In 6 weeks, how much do you think the price of 1 cup of local rice (sold loose) will be at the market?
    \item You answered with \{response to last question\} Why?
    \item What is the price of 1 liter of cooking oil in your market?
    \item In 6 weeks, how much do you think the price of 1 liter of cooking oil will be at the market?
    \item You answered with \{response to last question\} Why?
    \item What is the price of 1 kilo of beef (bone-in) in your market?
    \item In 6 weeks, how much do you think the price of 1 kilo of beef (bone-in) will be at the market?
    \item Would you agree to participate in future research with us?
    \item Please enter your email address below for future opportunities to receive rewards for participating in other surveys.
    \item Please provide your phone number below to qualify for mobile data. Include the country code. (If you are not interested, leave this field blank and click "Submit").
\end{enumerate}
The survey in Hausa is available upon reasonable request.

\subsection*{Graphics for response time and length.}
\label{S3_Appendix}
We show all the histograms for the response time and length broken up by month and sex.  The data has a long tail, so for viewing the x-axis label `$+$' refers to responses over 17 min for response time and over 480 characters for response length. 

\begin{figure}[!ht]
\includegraphics[width=\textwidth]{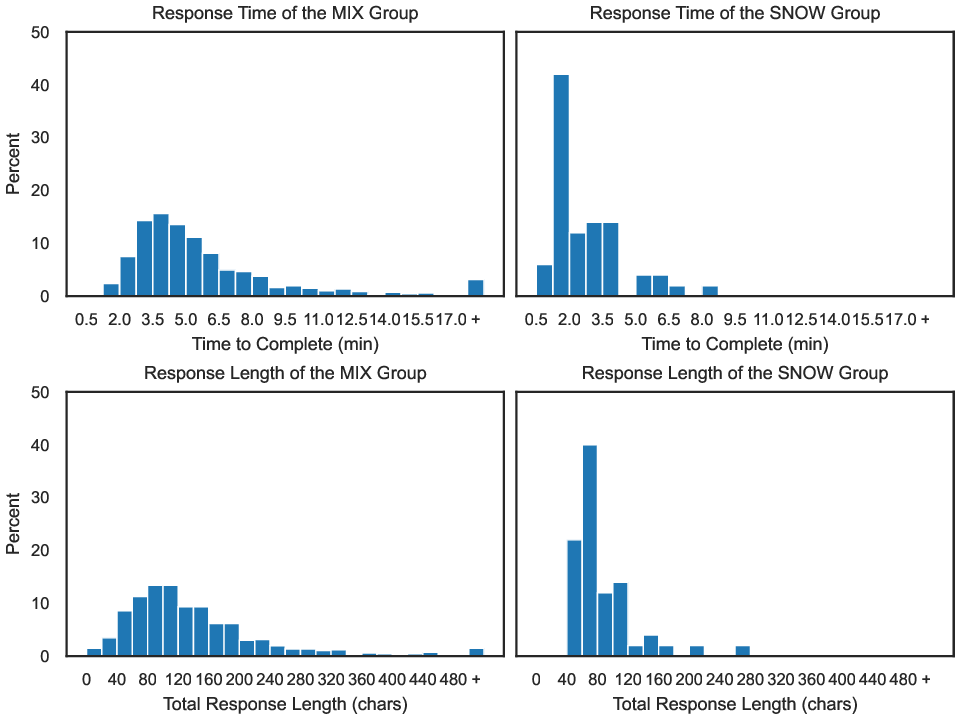}
\caption{{\bf Response time and length for men in the MIX and SNOW group in August}.}
\end{figure}

\begin{figure}[!ht]
\includegraphics[width=\textwidth]{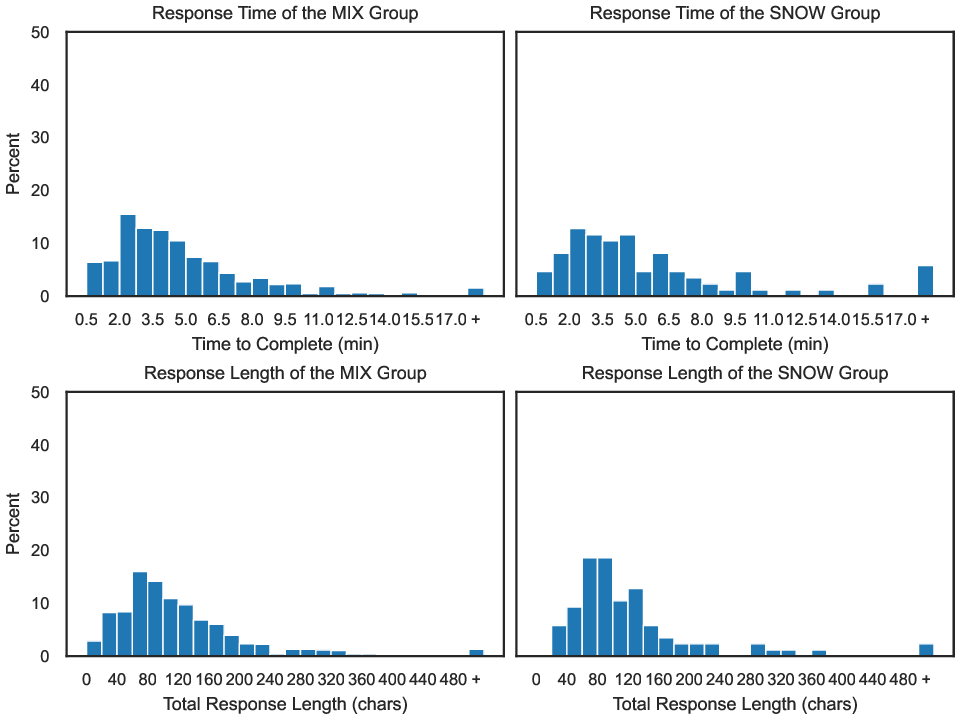}
\caption{{\bf Response time and length for men in the MIX and SNOW group in September}.}
\end{figure}

\begin{figure}[!ht]
\includegraphics[width=\textwidth]{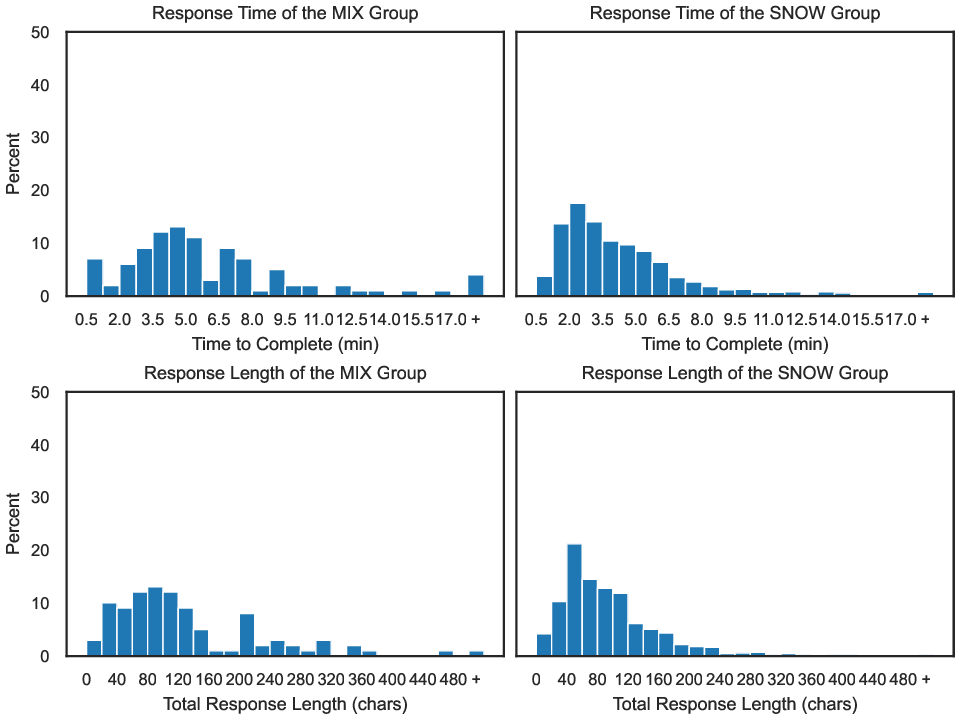}
\caption{{\bf Response time and length for men in the MIX and SNOW group in October}.}
\end{figure}

\begin{figure}[!ht]
\includegraphics[width=\textwidth]{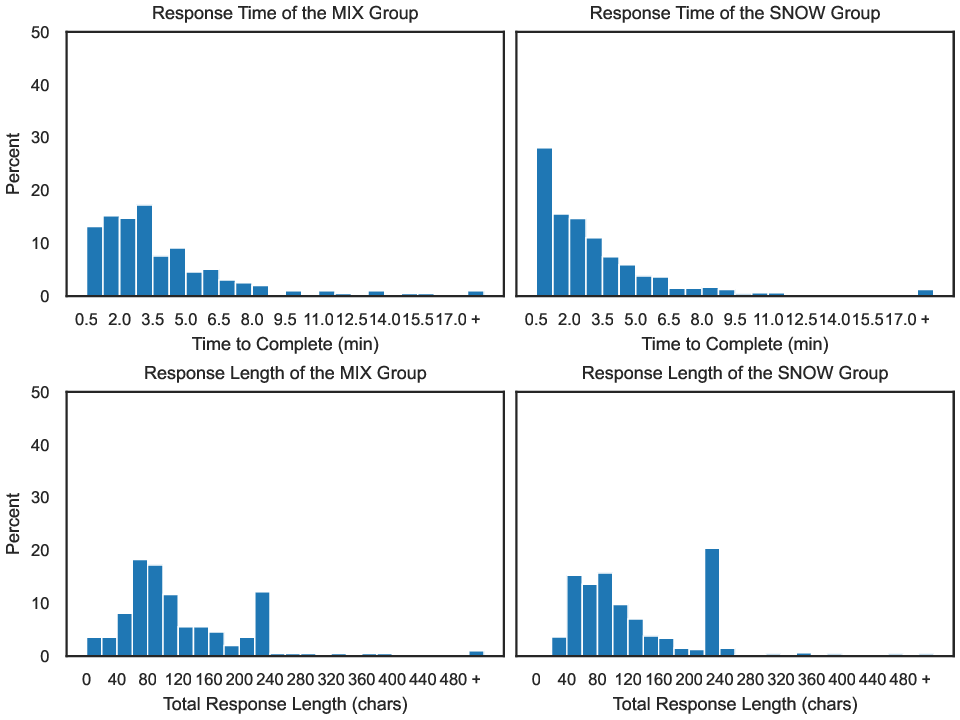}
\caption{{\bf Response time and length for men in the MIX and SNOW group in January}.}
\end{figure}

\begin{figure}[!ht]
\includegraphics[width=\textwidth]{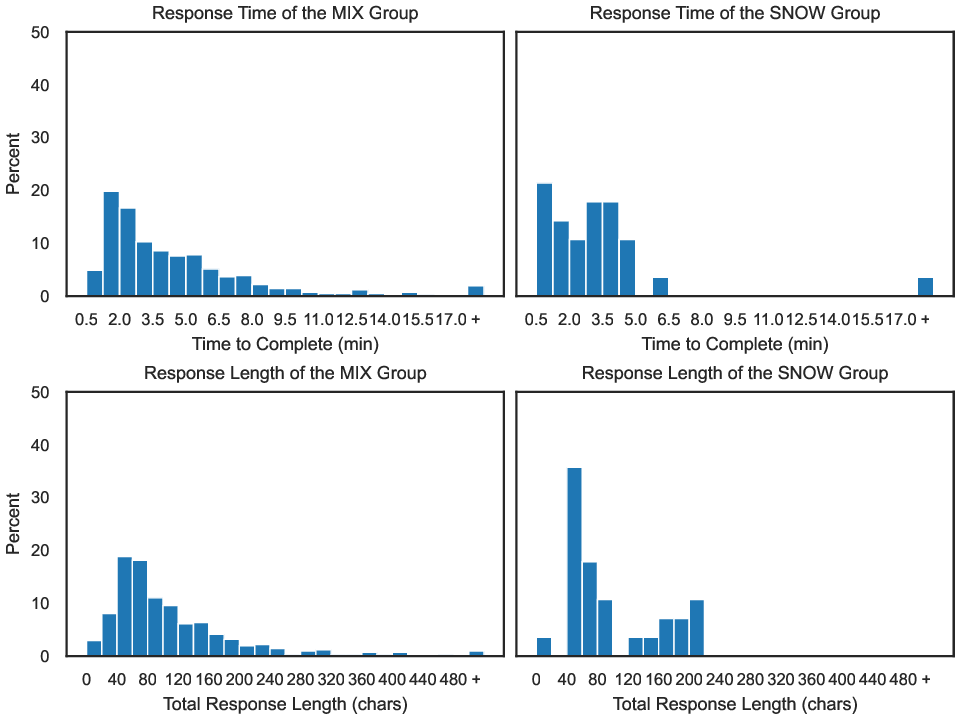}
\caption{{\bf Response time and length for men in the MIX and SNOW group in February}.}
\end{figure}

\begin{figure}[!ht]
\includegraphics[width=\textwidth]{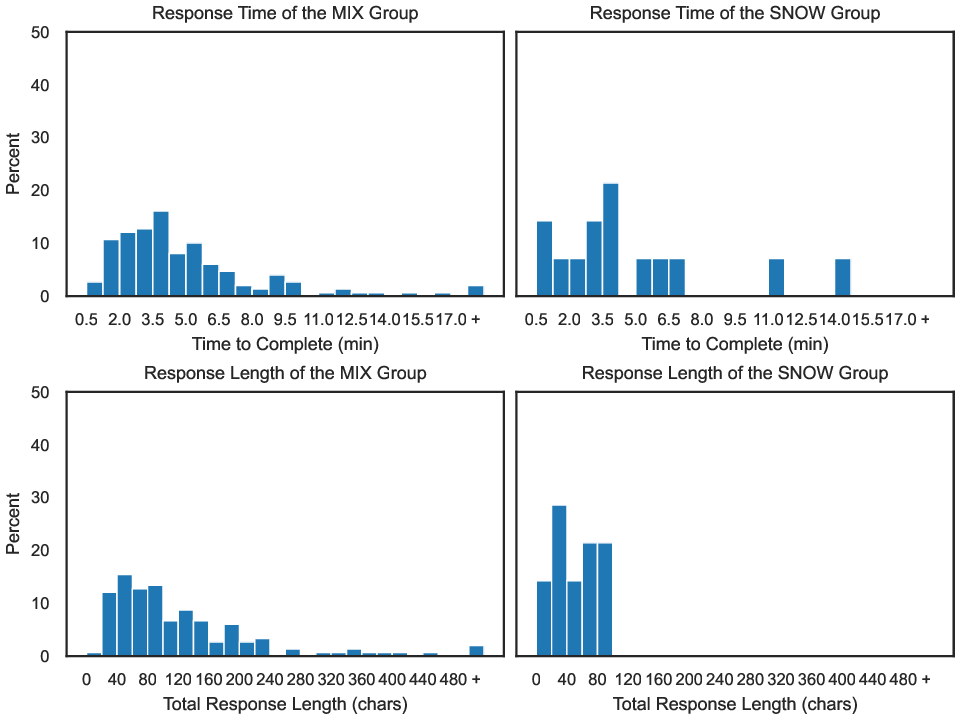}
\caption{{\bf Response time and length for men in the MIX and SNOW group in March}.}
\end{figure}

\begin{figure}[!ht]
\includegraphics[width=\textwidth]{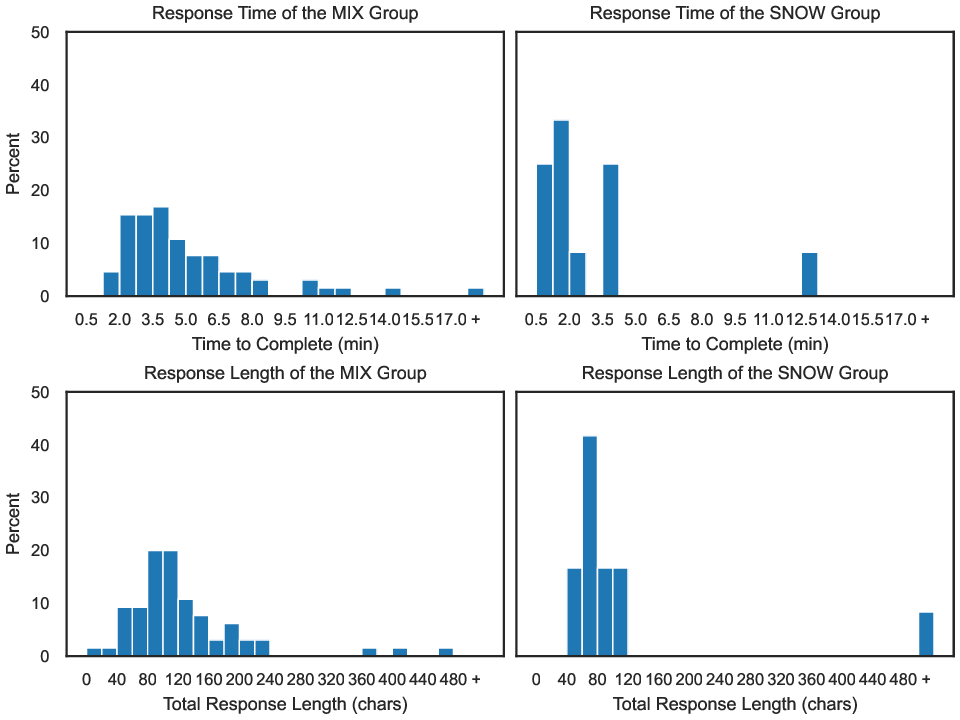}
\caption{{\bf Response time and length for women in the MIX and SNOW group in August}.}
\end{figure}

\begin{figure}[!ht]
\includegraphics[width=\textwidth]{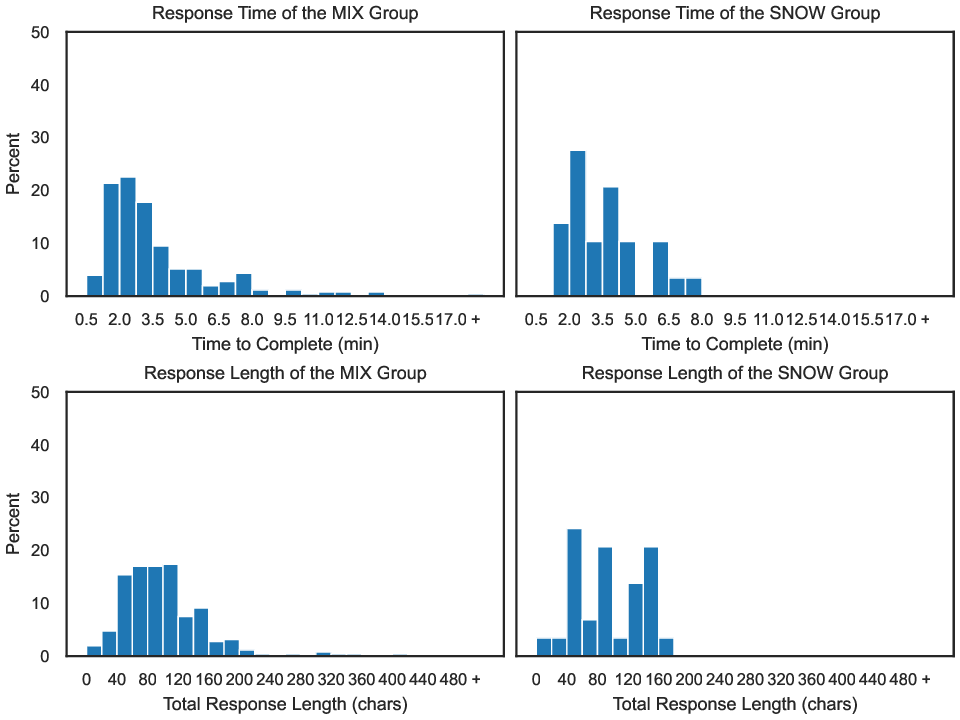}
\caption{{\bf Response time and length for women in the MIX and SNOW group in September}.}
\end{figure}

\begin{figure}[!ht]
\includegraphics[width=\textwidth]{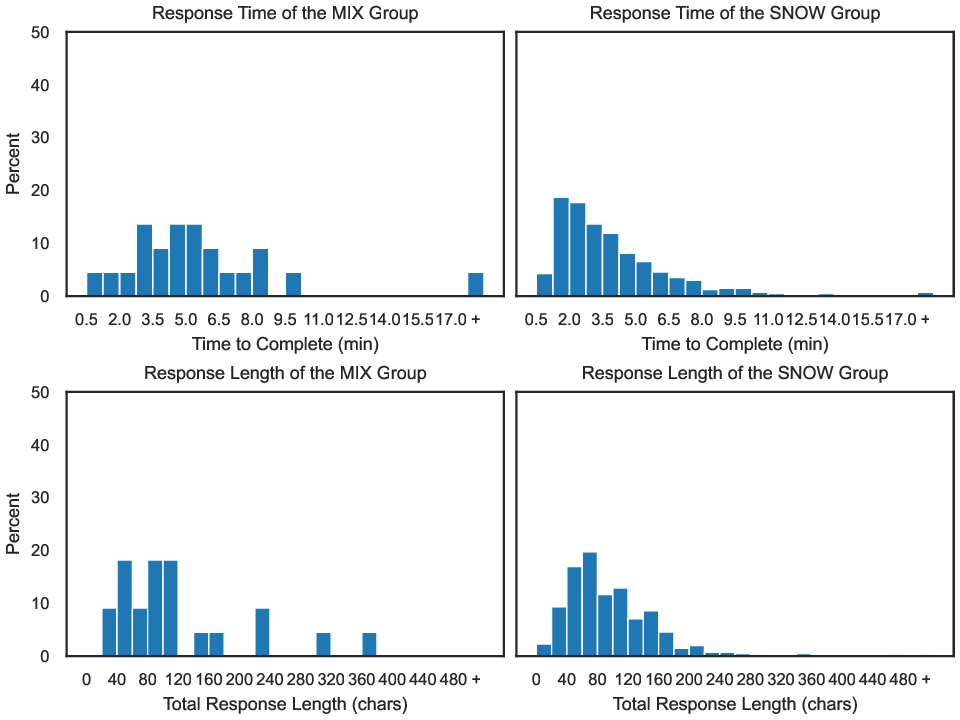}
\caption{{\bf Response time and length for women in the MIX and SNOW group in October}.}
\end{figure}

\begin{figure}[!ht]
\includegraphics[width=\textwidth]{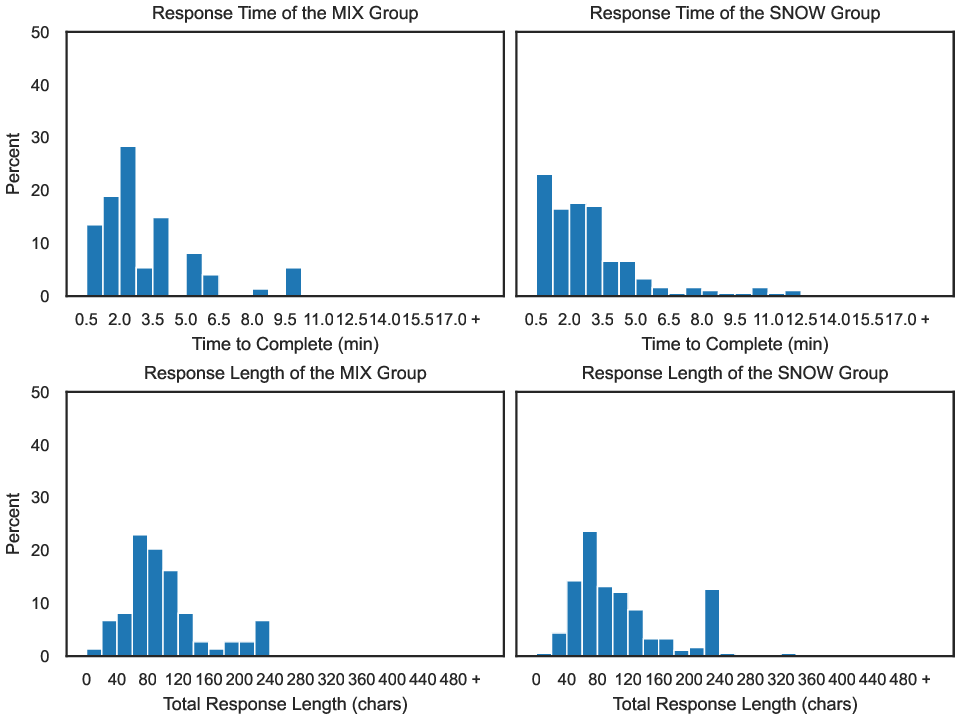}
\caption{{\bf Response time and length for women in the MIX and SNOW group in January}.}
\end{figure}

\begin{figure}[!ht]
\includegraphics[width=\textwidth]{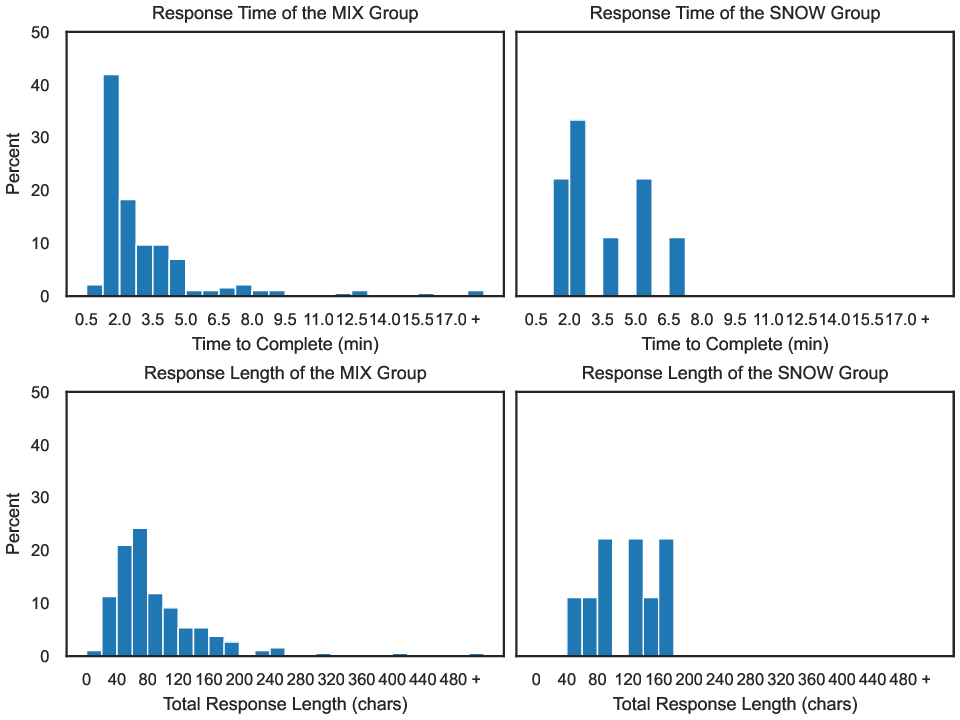}
\caption{{\bf Response time and length for women in the MIX and SNOW group in February}.}
\end{figure}

\begin{figure}[!ht]
\includegraphics[width=\textwidth]{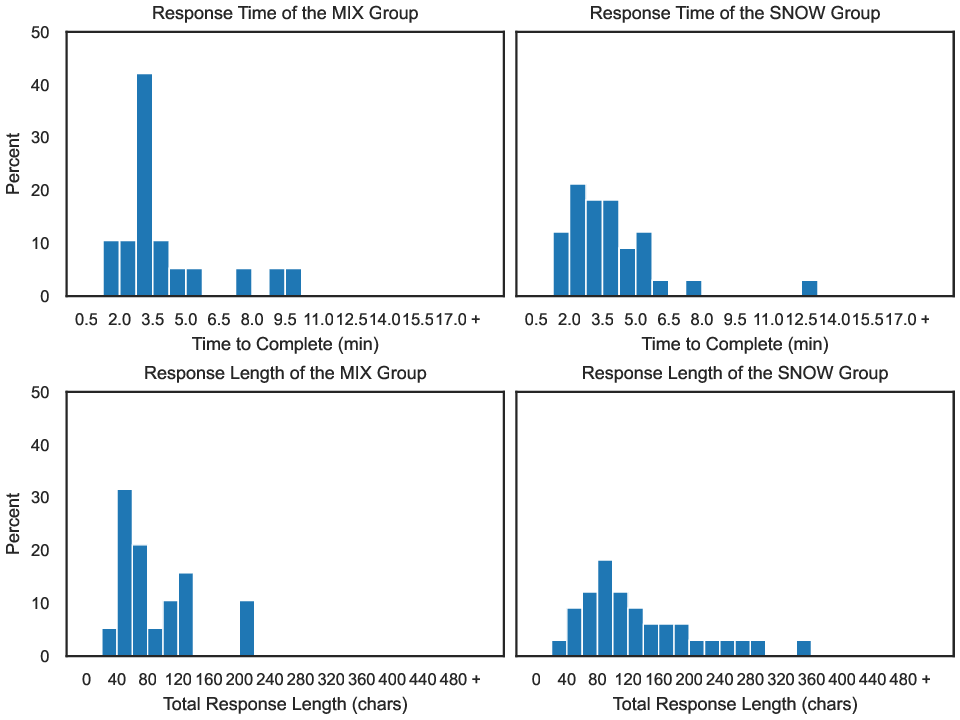}
\caption{{\bf Response time and length for women in the MIX and SNOW group in March}.}
\end{figure}

\end{document}